\documentclass[10pt,conference]{IEEEtran}
\IEEEoverridecommandlockouts

\usepackage[table,usenames,dvipsnames]{xcolor}
\usepackage[hidelinks]{hyperref}
\usepackage{lipsum}
\usepackage{framed}
\usepackage{multirow} 
\usepackage{enumitem}
\usepackage{soul}
\usepackage{amsmath}
\usepackage{amssymb}
\usepackage[para]{threeparttable}
\usepackage{mdframed}
\usepackage{float}
\usepackage{hanging}
\usepackage{amssymb}
\usepackage{csquotes}
\usepackage{wasysym} 

\usepackage[pdftex]{graphicx}
\usepackage{subcaption}
\graphicspath{{./fig/}}
\DeclareGraphicsExtensions{.png,PNG,.pdf,.jpg,.jpeg}

\usepackage{cite}

\usepackage{pifont}

\usepackage{tikz}

\definecolor{awhite}{RGB}{0, 0, 0}
\definecolor{steelblue}{RGB}{0, 0, 152}
\definecolor{royalblue}{RGB}{65,105,225}
\definecolor{rufous}{rgb}{0.66, 0.11, 0.03}

\definecolor{lightgray}{gray}{0.9}

\newcommand{\red}[1]{#1}

\newcounter{obs}
\stepcounter{obs}

\sethlcolor{lightgray}

\newcommand{\api}{\textit{SEFrame}}

\newcommand*\circled[1]{\tikz[baseline=(char.base)]{
            \node[shape=circle,draw,inner sep=2pt] (char) {#1};}}

\begin{document}

\title{Assessing Semantic Frames to Support Program
Comprehension Activities
\thanks{This work was supported under the Natural Sciences and Engineering Research Council of Canada (RGPIN-2016-03758).}}

\author{\IEEEauthorblockN{Arthur Marques, Giovanni Viviani, Gail C. Murphy}
\IEEEauthorblockA{
Department of Computer Science\\University of British Columbia, Canada\\
\{msarthur, vivianig, murphy\}@cs.ubc.ca}
}

\maketitle

\begin{abstract}
Software developers often rely on natural language text that appears
in software engineering artifacts to access critical information as
they build and work on software systems. For example, developers
access requirements documents to understand what to build,
comments in source code to understand
design decisions,
answers to questions on Q\&A sites to
understand APIs, and so on. To aid
software developers in accessing and using this natural language
information, software engineering researchers often use techniques
from natural language processing. In this paper, we explore whether
frame semantics, a general linguistic approach, which has
been used on requirements text, can also help address
problems that occur when applying lexicon analysis-based techniques to
text associated with
program comprehension activities. We assess the applicability of generic semantic frame
parsing for this purpose, and based on the results, we propose
\api{} to tailor semantic frame parsing for program comprehension
uses. We evaluate the correctness and robustness of the
approach finding that \api{} is correct in between 73\% and 74\% of
the cases and that it can parse text from a variety of software artifacts used to support program
comprehension.
We describe how this approach could be used to enhance
existing approaches to identify meaning on intention from software
engineering texts.

\end{abstract}


\section{Introduction}
\label{sec:introduction}

Much critical information about software is captured in natural
language. For instance, software requirements are often expressed in
natural language text, question-and-answer web sites that enable developers to help
each other rely on natural language explanations, and comments written
in natural language within
source code help a developer understand what a piece of code
is intended to do.

Researchers have long recognized the value of this natural language
text, utilizing various techniques to extract
information from this text that can be embedded in
tools for software developers.
A number of the techniques commonly employed by researchers are based on the
frequency of co-occurrence of words (or phrases) in documents. An
early example is Maarek and Smadja's use of lexical relations to index
software libraries~\cite{maarek1989}. Since this early use, software engineering
researchers have continued to leverage advances in
these approaches, such as when
Maletic and Marcus applied Latent Semantic
Analysis (LSA)~\cite{deerwester1990LSI} to help cluster software components to aid
program comprehension of a software system~\cite{Marcus2003}, or when Nguyen and colleagues
applied Word2Vec~\cite{mikolov2013word2vec} to support the retrieval of API
examples~\cite{nguyen2017}.

At times, software engineering researchers have argued
that general lexicon analysis techniques from natural language
processing are insufficient to address text appearing in
software engineering artifacts.
For example, Di Sorbo and colleagues
argued that lexicon analysis, like LDA, was insufficient to
classify emails based on developers'
intentions~\cite{DiSorbo2015}.
 Based on a manual
analysis of existing text in developer emails, they defined rules to
distinguish sentences discussing feature requests, asking for an
opinion, or proposing solutions, amongst others.  As another example,
Gu and Kim argued lexicon analysis was insufficient to
assess the intent of user reviews and, like Di Sorbo, proceeded
with a manual analysis~\cite{gu2015}.

These arguments that lexicon-based natural 
language techniques are not applicable is often based
on a need for access to the \textit{meaning} of
sentences in the natural language text~\cite{DiSorbo2015, gu2015, Arya2019}. Whenever similarity arises in the reasons for
why existing approaches are insufficient,
it raises the question
of whether there exists an approach that might fill
the gap.

In this paper, we consider whether frame
semantics~\cite{fillmore1976frame, Baker1998}---a general
linguistic approach---might be such a gap filler.  The frame semantics
theory is based on how readers comprehend the roles that words take in
a sentence with respect to events of interest~\cite{Baker1998,
  jurafskyspeech}. We provide a detailed
  explanation of frame semantics in the next
  section (Section~\ref{sec:background}).

To the best of our knowledge, there have been only a few uses of frame
semantics in software engineering research~\cite{jha2017, kundi2017, alhoshan2019using, Marques2020} (Section~\ref{sec:related-work}). These approaches
have largely focused on text associated
with software requirements, leaving open the
question of applicability of the approach to
text in documents supporting program
comprehension activities.
One paper~\cite{Marques2020} applies semantic
frames to such documents, but does so from the
viewpoint of finding patterns in text rather than
considering the applicability of the technique
to the text.
It is thus an open question whether semantic
frames can help identify the meaning of
software engineering text associated with helping
developers build and comprehend programs, such as
text in API documents and bug reports.

To determine whether frame semantics are applicable to
a broad range of software artifacts
used for program comprehension, we consider three questions:

\begin{enumerate}
\item \textit{Do generic frame parsing approaches apply to text used for program comprehension activities?}  We
  assessed whether generic frame parsing as provided by the SEMAFOR
  tool~\cite{das2014frame} applies to such text by having two annotators label 1,866 sentences
  sampled from 1,802 documents drawn from existing datasets
  (Section~\ref{sec:coding}). We found that 5 frames required
  modification to properly capture the meaning of the text from documents associated with program comprehension
  activities and that 10 other frames were not
  suitable. We introduce \api{} that supports this tailoring of
  frame parsing to program comprehension related software engineering artifacts (Section~\ref{sec:api}).
\item \textit{Does~\api{} produce correct frames?}
  We assessed whether our \api{} approach provides results
  acceptable to a broader range of evaluators by having 10 evaluators evaluate 360 sentences and one of their associated frames
  (Section~\ref{sec:study-I}). We found that the frames parsed by \api{} were identified as correct in 73\% of the cases, 
  \red{a number comparable to SEMAFOR's 75\% correctness score,} 
  indicating that our approach produces correct frames.
\item \textit{Is~\api{} robust?}
    We applied \api{} to an additional 5 types of software
    engineering artifacts and had an additional 20 evaluators
    assess the results as a means of further testing the
    generalizability of \api{} (Section~\ref{sec:study-II}). We found that the evaluators found 74\% of the frames
    to be correct, indicating that our approach is generalizable to a broader range  of artifacts
    associated with program comprehension.
\end{enumerate}

As a means of addressing the question of whether semantic frames might fill the gap for at least some techniques,
we discuss the application of semantic frames
to two existing software engineering techniques that rely on manual classification
(Section~\ref{sec:discussion}).
While our work provides a 
foundation from which to use semantic frames for software engineering, future
work should continue to extend the investigation of the applicability of less
commonly occurring frames.

This paper makes four contributions:
\begin{itemize}
\item It introduces a refined version of semantic frame parsing for program comprehension text called \api{}.
\item It demonstrates that \api{} produce correct and robust frames for text from software engineering artifacts associated with program comprehension
\item It presents labelled data of sentences extracted from a variety of software artifacts and their associated frames.
\item It provides \api{} as a Python API.
\end{itemize}

We begin by providing an example of
semantic frame parsing (Section~\ref{sec:background})
and reviewing related work (Section~\ref{sec:related-work}).
We then describe our  investigation of
the applicability of frame semantic to the program
comprehension related text (Section~\ref{sec:coding}), leading to the development of \api{} (Section~\ref{sec:api}). We describe our evaluation of \api{} (Sections~\ref{sec:study-I} and~\ref{sec:study-II}) and place those results in the context of threats
to their validity (Section~\ref{sec:threats}).
We discuss implications and final remarks in Sections~\ref{sec:discussion} and ~\ref{sec:conclusion}, respectively.

Our data is available at this link\footnote{\url{https://figshare.com/s/323a80bd801d45ed7b89}} and will be made publicly
available upon acceptance of this paper.

\section{Background}
\label{sec:background}
\begin{figure}[t]
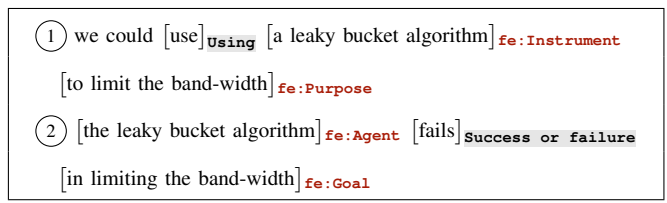


\smallskip
\begin{footnotesize}
\begingroup
\setlength{\tabcolsep}{10pt} \renewcommand{\arraystretch}{2} \begin{tabular}{|l|}
    
    \hline

    \circled{\scriptsize 1} 
    
    we could  $\big[$use$\big]$\textsubscript{\ttfamily \hl{\textbf{Using}}} $\big[$a leaky bucket algorithm$\big]$\textsubscript{\ttfamily \color{rufous} \textbf{fe:Instrument}} \\

    ~~~$\big[$to limit the band-width$\big]$\textsubscript{\ttfamily  \color{rufous} \textbf{fe:Purpose}}  \\
    
    \circled{\scriptsize 2} 
    
    $\big[$the leaky bucket algorithm$\big]$\textsubscript{\ttfamily \color{rufous} \textbf{fe:Agent}} 
    
    $\big[$fails$\big]$\textsubscript{\ttfamily \hl{\textbf{Success or failure}}} \\

    ~~~$\big[$in limiting the band-width$\big]$\textsubscript{\ttfamily \color{rufous} \textbf{fe:Goal}}   \\

    \hline
\end{tabular}
\endgroup
\end{footnotesize}
\caption{Example of frames and frame elements}
\label{fig:frame-example}
\end{figure}

Program comprehension activities often require 
inquiry of information available in text documents.
Within these documents, a reader extracts
information based on 
the roles that words take in a sentence with
respect to events of interest~\cite{Baker1998, jurafskyspeech}.
Frame semantics~\cite{fillmore1976frame} theory provides lens to explain such a process,
where a `\textit{frame}' is the key unit that assists understanding readers' comprehension.
The FrameNet project realizes frame semantics. It is
a large collection of more than
1,200 frames recognizable in the English language~\cite{baker1998berkeley}
and semantic role parsers~\cite{jurafskyspeech}
have leveraged annotated data available in the FrameNet project
for automatically identifying frames and their associated data~\cite{das2010, das2012, swayamdipta17}.
SEMAFOR is an example of a such a semantic frame parser~\cite{das2014frame}
and it uses a statistical model for determining 
which word tokens evoke frames from FrameNet.
SEMAFOR is also able to identify predicates not seen in the FrameNet lexicon,
what makes the tool particularly useful for identifying frames 
in domain-specific tasks (e.g.,~\cite{moschitti2003} or~\cite{Sogaard2015}).

To illustrate frame semantics in action, we utilize an example
from Di Sorbo et al.~\cite{DiSorbo2015}, whose approach aimed to classify the content of development emails in six categories, such as \textit{`solution proposal'} and \textit{`problem discovery`}. Let's consider the following two sentences:
\begin{itemize}
\item  \textit{we could use a leaky
  bucket algorithm to limit the band-width}; and
\item \textit{the leaky bucket algorithm fails in
  limiting the band-width}.
\end{itemize}
These two sentences are lexically similar, making it
difficult to use lexicon analysis techniques to categorize them. Di Sorbo's work categorized both sentences as \textit{`solution proposal`}, even though a more accurate label for the second one would have been \textit{`problem discovery'}.

If we apply frame semantics to these sentences, using SEMAFOR, the differences
become apparent. 
Figure~\ref{fig:frame-example} presents the results of a frame
analysis for the sentences.
The frames of each sentence (in grey) represent a
triggering event and the \textit{frame elements (fe)} (in red) are arguments needed
to understand the event. The enclosing square brackets
mark all lexical units, or words,
associated with either a frame or a frame element.

In the first sentence, the \textit{Using} frame 
captures that an \texttt{instrument}, the leaky bucket algorithm, is
manipulated to achieve a \texttt{purpose}, namely to limit band-width.
In contrast, in the second sentence, the \textit{Success or Failure}
frame identifies the entity, or \texttt{agent}, that fails to achieve the \texttt{goal} of limiting the band-width.
With the frame semantics it becomes possible to distinguish meanings of these
two lexically similar sentences in such a way that the categorization desired
by Di Sorbo et al. might occur. For instance, a categorization technique
might associate a \textit{Using} frame  with a category
such as \textit{`solution proposal'}.
\section{Related Work}
\label{sec:related-work}

Within software engineering, frame semantics have largely been applied to problems related
to requirements engineering.
Recognizing the limitations of approaches based on 
lexicon analysis, Jah and Mahmoud used frame semantics
to represent text in user reviews of applications
to improve the determination via a classifier of 
which reviews are feature requests, which are bug requests and which
are something else~\cite{jha2017}. They concluded that a
frame semantics
classifier based on SEMAFOR
reduced the chance of overfitting and led to better
performance over unseen data. Jah and Mahmoud applied frame semantics
without any assessment of how well the general frame parsing
approaches perform on software engineering text.

Kundi and Chitchyan used frame semantics to assist with requirements
elicitation~\cite{kundi2017}. Their work seeks to identify actors,
roles, and relationships according to the frames and frames elements
extracted from requirements specification text.  As part of this
research, they observed that certain frame elements do not reflect the
text from which they were extracted~\cite{kundi2017}.

Alhoshan et al. also applied frame semantics to problems
associated with software requirements.
Their investigation on 
the use of frame semantics in 18 requirements documents~\cite{alhoshan2018a, Alhoshan2018b}
has identified that only some of the frames from FrameNet (123 frames) relate to text in requirements engineering.
The identified frames were then used in a follow-up study~\cite{alhoshan2019using} 
for the design of a technique able to identify textual related requirements
using both Word2Vec~\cite{mikolov2013word2vec} and FrameNet.
Although their study suggests that frames improve measuring  
semantic relatedness for text in requirements engineering, 
their results are bound  to the small set of artifacts from which 
frames were extracted and they discuss the need to extend their corpus and evaluation~\cite{Alhoshan2018b}.

Marques et al. used SEMAFOR to investigate the relevance of
text in software artifacts relevant to completing
software engineering tasks~\cite{Marques2020}.
These artifacts include ones that support
program comprehension activities, such as
referring to bug report discussion to confirm a system's behaviour.
In this study, semantic frame parsing was used as one
technique to look for patterns in text identified as relevant by participants. 
A total of 346 distinct frames over 
20 software development natural language artifacts were identified and 
results suggest consistency in the frames of sentences considered relevant. This work did not investigate
whether the identified frames were meaningful
for the sentences parsed.

This paper extends these 
investigations 
by 
considering text from documents associated
with program comprehension activities,
by tailoring
frequently occurring frames in these artifacts to
appropriate meanings for
program comprehension, and
by investigating the correctness of frames parsed with
a larger set of evaluators.

\begin{table*}
\caption{Datasets for Annotation and Correctness Studies}

\begin{threeparttable}
\rowcolors{2}{}{lightgray}
\begin{tabular}{llllcc}
\hline    
\textbf{Dataset} & \textbf{Source} & \textbf{Description} & \textbf{Size} & \textbf{Sentences} & \textbf{Sample} \\ 
\hline
\hline
InfoTypes &
\cite{Arya2019} & 
\parbox[l][0.9cm][c]{9cm}{Annotated data of 15 complex issue threads across three projects hosted on GitHub} &
15 GitHub issues &
5981  &
597 
\\
AnswerBot &
\cite{Xu2017} & 
\parbox[l][0.9cm][c]{9cm}{Technical questions and answers from Stack Overflow covering a diversity of aspects of Java programming } &
100 SO questions &
3306  & 
552
\\
MissingInf &
\cite{Chaparro2017} & 
\parbox[l][0.9cm][c]{9cm}{Bug reports from nine software projects of different types and domains. Bug reports mined both from Bugzilla and GitHub} &
2,912 bug reports & 
44554  &
653
\\
APIPatterns\tnote{\dag} &
\cite{Maalej2013} & 
\parbox[l][0.9cm][c]{9cm}{Annotated data of knowledge patterns found in the JDK 6 and .NET 4.0 reference API } &
5,574 API elements &
 4451  &
577
\\
\hline
\end{tabular}
\begin{tablenotes}
    \item[\dag] \scriptsize Comprises only Java API documentation
\end{tablenotes}
\end{threeparttable}    
\smallskip
\label{tbl:datasets}
\end{table*}

\section{Do Generic Frame Parsing Approaches Apply to Program Comprehension?}
\label{sec:coding}

Ideally, frames  identified by semantic frame parsers would find appropriate meaning in
text related to software engineering artifacts `out-of-the-box' without change. As already seen
in the investigation by Alhosan et al. on a small number of software requirement documents~\cite{Alhoshan2018b}, this
assumption does not hold.

To explore how many frames might need alteration for software
engineering text, we ask: ``do generic frame parsing approaches apply to
text that appears in software engineering artifacts?''
Specifically, we 
investigate the frames that result when a standard
parser, SEMAFOR, is applied to 
a broader range of artifacts associated with
program comprehension activities.
We are particularly interested in how the presence of
jargon or text specific to software 
such as method signatures, stack traces, or command-line arguments,
might
affect the meaning of the frames identified.

To answer this question, we had two
annotators perform an open coding of frames parsed from three different 
datasets drawn from corpa published in previous software engineering studies.
Our open coding analysis comprises the inspection of the 50 most frequently occurring distinct frames 
that appear in 1,866 sentences. 

\begin{figure} 
\includegraphics[width=\linewidth]{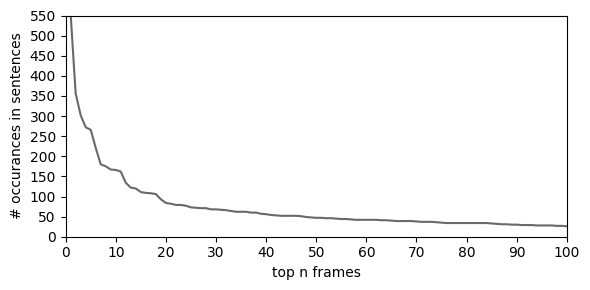}
\vspace{-0.5cm}
\caption{Distribution of frames over sampled sentences}
\label{fig:frame-distribution}
\end{figure}

\subsection{Method}

\subsubsection{Datasets}

For this investigation, we chose artifact types that relate to program comprehension tasks~\cite{Li2013, Ponzanelli2017}.
These artifact types have been the target
of many researchers given the prominent role they play in the activities of software developers:

\begin{itemize}
    \item developers seek Q\&A websites, such as Stack Overflow, to analyze and adapt solutions available online for their current development tasks~\cite{Zhang2019, Xia2017, umarji2008archetypal, Brandt2009a, Treude2011a}; and 
    \item bug reports serve as a central place for several program comprehension activities, such as documentation, coding, or testing~\cite{Rastkar2010, aranda2009secret, Lotufo2012}.
\end{itemize}

To minimize any bias in our selection of artifacts, we chose to rely on existing datasets published at
major software engineering conferences. For this investigation, we use the
first three  datasets listed in Table~\ref{tbl:datasets}.

\subsubsection{Sampling} 

As selected datasets comprise more than 53,800 sentences, 
and given that sentences can contain more than one frame, 
manually
inspecting the frames in each sentence is infeasible. We thus
sample a statistically significant number of sentence-frame pairs for inspection.

We follow the procedures described by Bacchelli and
colleagues~\cite{Bacchelli2010, Bacchelli2012}, statistically sampling
sentences from each dataset with a 90\% confidence level and a 5\%
error level with respect to the original number of sentences of that
dataset. As Table~\ref{tbl:datasets} shows, sample sizes 
range from 552 to 653 sentences from each source.
We then use SEMAFOR~\cite{das2014frame} to semantically parse each
sentence and obtain its set of frames, i.e.,
when SEMAFOR is applied to a sentence, it may
identify multiple frames for the sentence.
We chose to use SEMAFOR because
previous studies in the field have used the tool and adopting the same
parser facilitates comparison (Section~\ref{sec:related-work}).

After applying SEMAFOR
to the dataset, we identify 444 distinct frames.
\red{Figure~\ref{fig:frame-distribution} plots the distribution of 
\textit{all} sentences per frame.}
The distribution follows Zipf's law with certain frames appearing in a high number of sentences and a long tail where some frames are specific to a few sentences.
\red{The most occurring frames appear in 30\% (topmost frame) to 10\% (top 10th frame) of the sampled sentences.
In comparison, the 100th frame appears in no more than 25 sentences, which accounts for 1\% of the sentences sampled.}
We are most interested in those frames that occur
frequently and so, 
we produce the final 
set of sentence-frame pairs for inspection by sampling sentences for each of the top 50 frames.
That is, given all sentences associated to a frame, select a statistically significant number of sentences for inspection.

Our sampling procedures resulted in a set of 1,866 sentence-frame pairs 
for inspection that comprise all occurrences of the top 50
distinct occurring frames.

\subsubsection{Annotation Process}

Similar to Alhoshan and colleagues~\cite{alhoshan2018a}, we use an
annotation approach which consists of inspecting the frame, its
definition according to the FrameNet
database\footnote{https://framenet.icsi.berkeley.edu/fndrupal/frameIndex},
and its frame elements and associated lexical units.  For each frame,
annotators indicated if the frame was correct  and, if not,
suggested
either a replacement from FrameNet or an entirely new frame.

As a first step to get familiarized with the definition of each frame encountered,
the annotators read a total of 180 pairs of sentences with their associated parsed frame.
Then, the two annotators (both authors of the paper) proceeded iteratively.
In the first iteration, 300 sentence-frame pairs were individually annotated by the two authors. 
At the end of this iteration, the 
two annotators discussed potential strategies for the resolution of
disagreed frames, resulting in an agreed output schema for the final iteration:
 a frame was annotated as \texttt{valid}, \texttt{invalid}, or
requires \texttt{modification} (with suggestions for a replacement).
At this point, the annotators individually annotated the final set of 1,866 sentence-frames.
We did not exclude the previous 300 pairs from this set, as 
returning to them ensures consistency in how conflicts were resolved.

\subsection{Results}

Table~\ref{tbl:annotation-results} summarizes results according to annotation outcomes: \texttt{valid}, \texttt{invalid}, or \texttt{modify}.
To assist discussion, we refer to the examples from Table~\ref{tbl:annotation-examples} by their unique identifiers (UID).

We found that the annotators agreed that SEMAFOR's parsing of the semantic frames was valid for 35 (64\%)
of the frames inspected.
An annotation of valid means that the meaning of frames
is  similar in both
every-day English text and software engineering text.  
In valid frames, SEMAFOR is also
able to correctly identify the predicates, namely the frame elements (in
red) and lexical units (enclosing brackets) of software engineering specific text.
For example, it successfully identifies
`turning off \texttt{XBitHack}' as the \textit{cause} for the
observed behaviour in $\text{UID}_{436}$ (Table~\ref{tbl:annotation-examples}).
For the \textit{Using} frame ($\text{UID}_{122}$), it also identifies the
`\texttt{JNI} function' as the instrument needed for achieving the
purpose of creating a certain class.

It is worth noting that even if a frame is considered valid
overall, some instances of where that frame is parsed may be invalid.
$\text{UID}_{120}$ in Table~\ref{tbl:annotation-examples} is an example of an invalid instance for the 
\textit{Using} frame, where `\textit{application}' does not evoke usage. 
Occurrences of invalid instances represent a small fraction of the data annotated frames and
we attribute invalid instances to misclassification errors inherent to SEMAFOR\footnote{SEMAFOR has an average precision of 75.54\% for the FrameNet
lexicon}.

\begin{table}
\centering
\caption{Summary of frames and annotation outcome with percentage of sentences containing the frames in the dataset}
\label{tbl:annotation-results}
\rowcolors{2}{}{lightgray}
\vspace{-1.5mm}
{\renewcommand{\arraystretch}{1.2}
\begin{tabular}{lll}

\hline
\textbf{35 frames (64\%)}
& 
\multicolumn{2}{l}{ \textbf{Valid}} 
\\
\hline

Intentionally act  & Quantity & Using \\ 
Temporal collocation & Being obligated  & Capability \\
Cardinal numbers  & Causation & Likelihood \\

Locative relation & Increment     &  Relational quantity \\

Point of dispute &  Existence & Intentionally create \\
Attempt & Instance &  Frequency \\
Desiring & Predicament &  Desirability \\
  
Required event &  Event & Identicality\\
Age & Gizmo &  Similarity  \\
  Time vector & Relative time &  Awareness \\
Measure duration &  Scrutiny & Sole Instance \\
Possession & Grasp & \\

\hline
\cellcolor{white} \textbf{5 frames (16\%)} 
& 
\multicolumn{2}{l}{\cellcolor{white} \textbf{Modified to} \textit{Execution}} 
\\
\hline

Arriving &  Means & Aggregate \\
Request &  Leadership & \\

\hline
\cellcolor{white} \textbf{10 frames (20\%)} 
& 
\multicolumn{2}{l}{\cellcolor{white} \textbf{Invalid}} 
\\
\hline
Statement & Type & Placing \\
Being named & Purpose & Roadways \\
Contingency  & Connectors & Text   \\
\hline
\end{tabular}
}
\end{table}

A total of 5 (16\%) frames required modification.
Annotators observed that the meaning of verbs such as `\textit{get}' or `\textit{run}' 
diverge from FrameNet's meaning as identified by SEMAFOR.
$\text{UID}_{1181}$ illustrates a frame that requires modification.
The verb `\textit{run}' evokes the \textit{Leadership} frame which 
is defined as control by a leader over a particular entity or group.
However, the sentence related to this frame in 
Table~\ref{tbl:annotation-examples} conveys executing an evaluation benchmark. 
Both annotators suggested  
 renaming the frame to \textit{Execution}
so that it reflects a meaning appropriate to the software engineering domain. 
The same logic applies to the \textit{Arriving} frame in $\text{UID}_{611}$.

\begin{table*}[]
\caption{Excerpt of results from annotation procedures}

\begin{threeparttable}    
\rowcolors{2}{}{lightgray}
\begin{tabular}{cllc}
\hline    
\textbf{UID} & \textbf{Frame} & \textbf{Sentence} & \textbf{Annotation} \\ 
\hline
\hline
436 &
Causation &
\parbox[l][0.7cm][c]{12.5cm}{
$\big[$Turning off XBitHack in my config$\big]$\textsubscript{\ttfamily \color{rufous} \textbf{fe:Cause}} 
$\big[$made$\big]$\textsubscript{\ttfamily \hl{\textbf{Causation}}} 
$\big[$this behavior go away$\big]$\textsubscript{\ttfamily \color{rufous} \textbf{fe:Effect}}    
} &
Valid
\\
122 &
Using &
\parbox[l][0.95cm][c]{12.5cm}{
I'm trying to $\big[$use$\big]$\textsubscript{\ttfamily \hl{\textbf{Using}}} 
$\big[$a JNI function$\big]$\textsubscript{\ttfamily \color{rufous} \textbf{fe:Instrument}} 
$\big[$to create a Java class$\big]$\textsubscript{\ttfamily \color{rufous} \textbf{fe:Purpose}} and set some properties of that class using the DeviceIdjava constructor method 
} &
Valid
\\
1181 &
Leadership &
\parbox[l][0.7cm][c]{12.5cm}{
Does anyone want to $\big[$run$\big]$\textsubscript{\ttfamily \hl{\textbf{Leadership}}}
$\big[$a benchmark?$\big]$\textsubscript{\ttfamily \color{rufous} \textbf{fe:Governed}}
} &
Modify
\\
611 & 
Arriving &
\parbox[l][0.8cm][c]{12.5cm}{
I cant even run this simple tensorflow script, as its result, I $\big[$get$\big]$\textsubscript{\ttfamily \hl{\textbf{Arriving}}}
$\big[$ImportError: No module named tensorflowpythonclient $\big]$\textsubscript{\ttfamily \color{rufous} \textbf{fe:Goal}}
} &
Modify
\\
120 &
Using &
\parbox[l][0.7cm][c]{12.5cm}{
Its a desktop standalone Java $\big[$application$\big]$\textsubscript{\ttfamily \hl{\textbf{Using}}} 
} &
Invalid
\\
1176 &
Roadways &
\parbox[l][0.7cm][c]{12.5cm}{
The $\big[$command line$\big]$\textsubscript{\ttfamily \hl{\textbf{Roadways}}}
is what almost every other application will use to build your JAR file    
} &
Invalid
\\

1493 &
Connectors & \parbox[l][0.8cm][c]{12.5cm}{
This means that at least the string-to-int mapping will stay consistent, even if $\big[$strings$\big]$\textsubscript{\ttfamily \hl{\textbf{Connectors}}} are passing out of memory
} &
Invalid
\\
263 &
\parbox[l][0.8cm][c]{1.5cm}{Being \\ obligated} &
\parbox[l][0.8cm][c]{12.5cm}{
$\big[$\texttt{I}$\big]$\textsubscript{\ttfamily \color{rufous} \textbf{fe:Responsible\_party}}
$\big[$have$\big]$\textsubscript{\ttfamily \hl{\textbf{Being\_obligated}}} 
two classes in a parent-child relationship
} &
Disagreement
\\
692 &
\parbox[l][0.7cm][c]{1.5cm}{Type} &
\parbox[l][0.7cm][c]{12.5cm}{
$\big[$\texttt{MQ}$\big]$\textsubscript{\ttfamily \color{rufous} \textbf{fe:Category}}
$\big[$version$\big]$\textsubscript{\ttfamily \hl{\textbf{Type}}} 71 on the server
} &
Disagreement
\\
\hline
\end{tabular}
\end{threeparttable}    
\smallskip
\label{tbl:annotation-examples}
\end{table*}

The 10 frames that were annotated as invalid account for 20\% of the inspected data. 
Annotators found that the presence of source-code related 
nouns is a common element to the invalid frames. 
\textit{Roadways} and \textit{Connectors} 
are examples of invalid frames. 
The sentences for these frames, as shown in Table~\ref{tbl:annotation-examples},
indicate that the nouns that evoke these frames are code-related, such as
`\textit{command line}' and `\textit{string}', and the frames parsed by SEMAFOR 
are not suitable for such cases.

Disagreements occurred in scenarios where one annotator suggested removal of a frame while the other said otherwise. 
As an example, the annotators disagreed on the resolution for
the \textit{Being obligated} frame
(Table~\ref{tbl:annotation-examples}). One annotator indicated
that \textit{Possession} would be a more appropriate frame, while the
other indicated that given that only a few sentences with the frame were invalid,
the situation is similar to a misclassification error inherent to using the SEMAFOR tool.
For other cases, such as in \textit{Type},
annotators observed that a single lexical unit was not enough to justify 
a new frame and thus, annotators agreed that the frame was invalid.

Out of the top 50 occurring frames detailed in
Table~\ref{tbl:annotation-results}, 35 (64\%) of them apply to text
from software engineering without modifications, 10 (20\%) are invalid and
16\% require modification\footnote{Detailed results are available in our \red{replication package}.}.

\section{\api: Tailoring FrameNet for Software Engineering}
\label{sec:api}

Our initial investigation found that modifications are needed to
SEMAFOR, and some of the frames defined in FrameNet, 
 to support appropriate parsing for software engineering
text. We describe~\api{}, a tool we have built that tailors SEMAFOR
for 
software design and other text related to program
comprehension activities.

We started the design of \api{} by gathering all frames that required modification 
or that are invalid and their respective lexical units. 
Then, we sought to identify patterns in the lexical units from these frames 
that would produce a simple set of heuristics to be used by \api{} 
to refine the frames extracted by SEMAFOR.

The first pattern observed by the annotators is that if the lexical unit of a frame is a verb with a particular meaning in the software engineering domain (e.g., \textit{run, call,} etc), the frame requires modification. 
A second pattern arises from frames evoked by source-code related nouns, which are deemed as invalid.
\red{Table~\ref{tbl:annotation-examples} provides examples for such patterns
where the verb `\textit{get}' in $UID_{611}$ 
and the word `\textit{string}' in $UID_{1493}$ 
are not correct for the \textit{Arriving} and the \textit{Connectors} frames, respectively.}

\red{The new proposed frame---Execution---addresses verb related modifications.}
\api{} makes verbs such as \textit{get, return, call, request, run} and \textit{process}
evoke the new frame instead. 
Interestingly, this change applies to all the frames
that require modification in Table~\ref{tbl:annotation-results}
since all of their verbs evoke executing system calls or running some procedure.

To address the second identified pattern,
\api{} removes any frames that are marked as invalid in Table~\ref{tbl:annotation-results}. 
Annotators identified that 
invalid frames are often related to code-related nouns
what could lead to the creation of a reference list for ignored lexical units.
Creating such a list poses the question of where code elements come from and whether the list is complete.
Therefore, \api{} opts to simple discard the frame 
rather than referring to which specific nouns and code-related terms would trigger removal.
We discuss this and other design decisions threats in Section~\ref{sec:threats}.

Intuitively, \api{} can be seen as a Decorator~\cite{gamma1995}
that extends the functionalities of the SEMAFOR tool. 
Its current version is written in Python and all its dependency 
libraries are provided via Docker containers~\cite{merkel2014docker}.

\section{Does~\api{} Produce Correct Frames?}
\label{sec:study-I}

For~\api{} to be usable by software engineering researchers, it must be seen
as parsing frames that are meaningful---correct---by more than those who
defined~\api{} (i.e., the two annotators). To explore if the results produced
by~\api{} are meaningful, we asked 10 evaluators, distinct from the
annotators in the initial investigation (Section~\ref{sec:coding}), to assess the
correctness of semantic frames parsed from sentences selected for all 
datasets shown in Table~\ref{tbl:datasets}.

\subsection{Method}

We applied \api{} to the datasets listed in Table~\ref{tbl:datasets}.
To provide a more generalizable result, we extended the
first three datasets used to create \api{} with a fourth dataset, 
\texttt{APIPatterns}, which has text extracted from Java
documentation (Table~\ref{tbl:datasets}). We include a new dataset as a first step towards investigating if and how frames apply to other types of sentences and also because API documentation is a resource that developers commonly 
refer to as part of a
program comprehension task~\cite{Robillard2011, lutters2007, lutters2007}.
For this study, we considered the 36 frames that are included in \api{}. 

The 10 evaluators were all currently graduate
students in an English-language based Computer Science program and
most of them had at least some previous industry experienced, averaging around 1.5 years. All evaluators were remunerated for their time spent on this work.

For each frame, we sampled 10 sentences that contained the specific
frame, for a total of 360 sentences. These sentences were divided in
10 batches, each one containing one sentence for each frame.
We then assigned three evaluators to each batch. We assigned evaluators with an approach based on similar studies~\cite{Viviani2019}: any group of three evaluators is evaluating maximum one
batch together and any group of two evaluators is evaluating
a maximum of two batches together.

\begin{figure*}
\centering
  \includegraphics[width=0.95\linewidth]{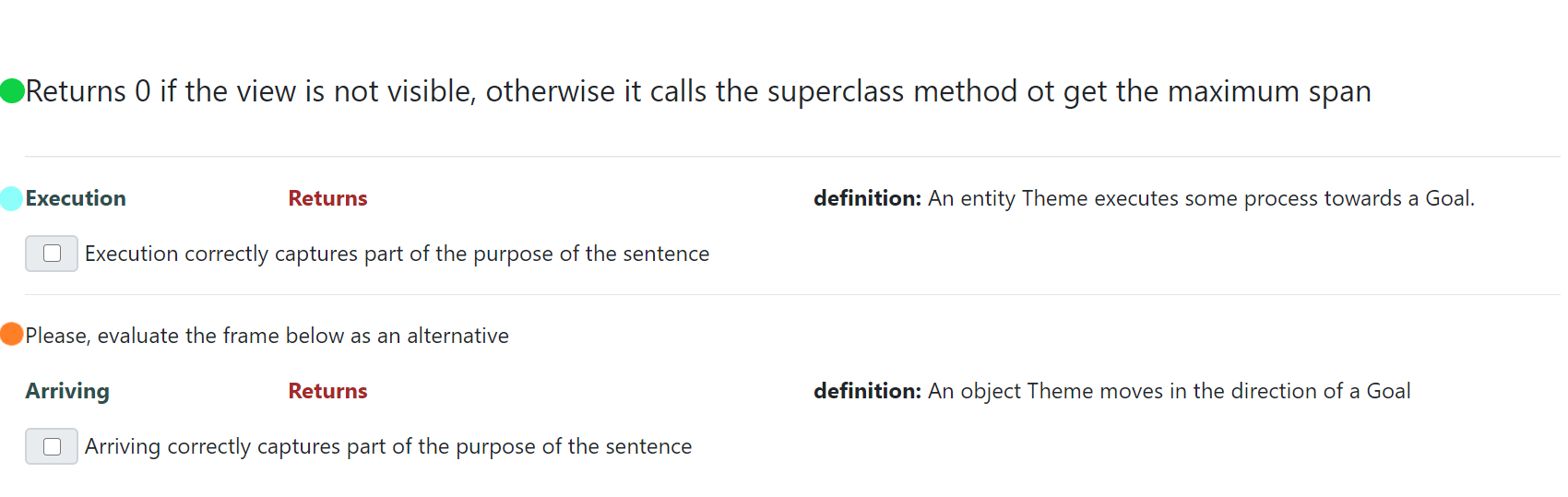}
  \caption{Example of a sentence and the associated frame in our online tool. The first line, next to the green dot, contains the sentence being evaluated. The next line, next to the light blue dot, contains the frame
  proposed by \api{}: from left to right, the name of the frame, the word that was used to identify the frame, and the definition of the frame; right below, the checkmark the evaluator could use to mark the frame as correct.
  The last line, next to the orange dot, contains the original frame proposed by SEMAFOR, in a similar fashion as the light blue line.}
  \label{fig:study1}
\end{figure*}

We used an online tool to present each sentence.
Each sentence and the associated frame was presented to the evaluator
through an online tool, an example of which can be seen in Figure~\ref{fig:study1}.

For each sentence, the evaluator was asked to indicate whether the associated
frame correctly represented the intention behind the words used in the sentence or not.
Whenever the evaluator indicated a frame to not be correct, a follow-up question displayed the
original frame parsed by SEMAFOR for that sentence and asked whether
the frame was more suitable. The evaluators shared a common chatroom with the authors where
some clarificatory questions were exchanged before the start of the study.
This common chatroom meant every evaluator had access to the same information.

\subsection{Results}
Table~\ref{tab:study1} presents a summary of the results with the percentage of frames that were evaluated as correct.  We define a frame to be correct if at least 2 out of the
3 evaluators agreed that it represents the meaning of it's associated sentence.
Overall, for 73\% of the 360 sentences were identified to be useful. The frame \textit{Execution}, that we introduced in \api{}, was identified as correct in 8 of the 10 sentences it appeared in. In the two sentences where it was considered incorrect, evaluators agreed that the original SEMAFOR frame was not a better candidate.
We consider this result evidence
that \api{} is able to detect correct frames.
\begin{table}[!h]
\centering
\caption{Results of Correctness Study}
\label{tab:study1}
{\renewcommand{\arraystretch}{1.05}
\rowcolors{2}{}{lightgray}
\begin{tabular}{lrlr}
\hline
\textbf{Frame} & \textbf{Ratio} & \textbf{Frame} & \textbf{Ratio} \\
\hline
\hline
Predicament & 100\% & Being obligated & 70\%\\
Required event & 100\% & Ordinal numbers & 70\%\\
Attempt & 100\% & Temporal collocation & 70\%\\
Identicality & 90\% & Measure duration & 70\%\\
Awareness & 90\% & Age & 70\%\\
Aggregate & 90\% & Quantity & 70\%\\
Likelihood & 90\% & Sole instance & 70\%\\
Existence & 90\% & Point of Dispute & 70\%\\
Desiring & 90\% & Relative time & 60\%\\
Instance & 80\% & Similarity & 60\%\\
Scrutiny & 80\% & Frequency & 60\%\\
Using & 80\% & Relational quantity & 60\%\\
Intentionally create & 80\% & Giving & 50\%\\
Capability & 80\% & Possession & 50\%\\
Grasp & 80\% & Time vector & 50\%\\
Inclusion & 80\% & Causation & 50\%\\
Desirability & 80\% & Cardinal numbers & 50\%\\
Execution & 80\% & Locative relation & 30\%\\
\hline
\end{tabular}
}
\end{table}

\section{Is~\api{} Robust?}
\label{sec:study-II}
Having shown in the previous section that~\api{} provides correct results in the view of evaluators independent from the creators of the tool, we turn to the question
of whether~\api{} can produce correct results over a broader range of kinds of artifacts. To explore if~\api{} generalizes across artifacts, we extend our dataset to cover an additional five types
of software artifacts---pull requests, vulnerability reports, a broader set of questions
and answers, mailing lists and app reviews---and evaluate the correctness of frames parsed by
individuals unknown to the \api{} creators.
\subsection{Method}

We extend the dataset 
with  sentences gathered from five different data sources,
which are associated with software design and programming: 

\begin{enumerate}
  \item \texttt{Pull requests} from the top 5 most starred projects on
    Github. From each project, we selected the top 10 most commented
    pull requests, and randomly sampled 50 comments for analysis.  We
    filtered comments with less than 50 characters to avoid common
    comments found in pull requests, e.g., ``Approved'' or ``Looks
    good to me''.
  
  \item \texttt{Security threats} related to vulnerability management
    data that are described using the Security Content Automation
    Protocol (SCAP)~\cite{waltermire2016}. For this dataset, we
    randomly select the description field of threat reports from 2019.
  
  \item A new set of questions from \texttt{Stack Overflow} due to the
    fact that the AnswerBot dataset used in our first study
    (Section~\ref{sec:study-I}) concerns only certain topics on the
    Java programming language~\cite{Xu2017}. For this new set of
    questions, we selected the 10 most commented questions from each
    one of the 5 programming languages mostly discussed on the
    platform.
  
  \item Developers' \texttt{Mailing lists} as represented by randomly
    selecting 6 archive threads from established projects of the
    Apache foundation.  From each thread, we parsed individual emails
    while ignoring any text in quote blocks such that we ignore
    duplicated text.
  
  \item \texttt{App reviews} from the 10 most popular apps from the
    Google Play Store. From each app, we selected the top 50 reviews
    and randomly sampled 25 reviews for analysis.
\end{enumerate}

\begin{table*}[]
\caption{Datasets - Robustness Study}
\begin{small}

\begin{threeparttable}    
\rowcolors{2}{}{lightgray}
\begin{tabular}{lllcc}
\hline    
\textbf{Dataset} & \textbf{Description} & \textbf{Size} & \textbf{Sentences} & \textbf{Sample} \\ 
\hline
\hline
\parbox[l][1.0cm][c]{2.0cm}{Pull Requests} &
\parbox[l][1.2cm][c]{9.5cm}{Pull requests from some of the most starred projects available on Github, i.e.,
Twitter's Bootstrap, Google's Flutter, Facebook's React, FreeCodeCamp, and OhMyZsh} &
25 pull requests &
648  &
328 
\\
\parbox[l][1.0cm][c]{2.0cm}{Security Vulnerability} &
\parbox[l][1.0cm][c]{9.5cm}{Security vulnerabilities available on the U.S. National Vulnerability Database} &
1866 reports &
4553  & 
579
\\
\parbox[l][1.0cm][c]{2.0cm}{Stack Overflow} &
\parbox[l][1cm][c]{9.5cm}{Stack Overflow most upvoted questions for 5 of the most popular programming languages available, i.e., javascript, java, python, c\#, and php.} &
50 SO questions & 
14196  &
633
\\
Mailing lists &
\parbox[l][1.2cm][c]{9.5cm}{Mail archives from popular projects hosted by the Apache foundation, namely, Apache Commons, Couch DB, HTTPD, Maven, Lucene, and Spark.} &
6 mail archives &
3650  &
561
\\
\parbox[l][1.0cm][c]{2.0cm}{Android \\ App Reviews} &
\parbox[l][1.0cm][c]{9.5cm}{User reviews for the ten most popular Android applications available on Google play store.} &
250 user reviews &
1212  &
429
\\
\hline
\end{tabular}
\end{threeparttable}    
\end{small}
\smallskip
\label{tbl:datasets-2}
\end{table*}

Table~\ref{tbl:datasets-2} details the sources used.
Overall, our selection criteria led to a total of 2530 sentences
containing 302 distinct frames.
We chose to focus on the top 20 most occurring \api{} frames as, similar to our initial
investigation in Section~\ref{sec:coding}, these frames appear in 33\% of the sentences in this new dataset.
We therefore selected the top 20 frames and randomly sampled one sentence for each.

\begin{figure}
\begin{subfigure}{.5\textwidth}
  \centering
  \includegraphics[width=.9\linewidth]{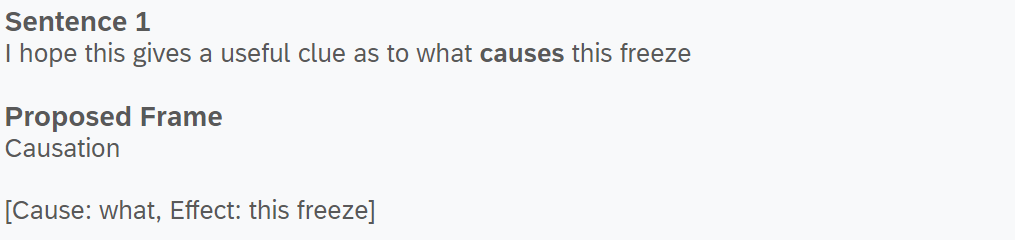}
  \caption{Correct frame}
  \label{fig:correct}
\end{subfigure}
\begin{subfigure}{.5\textwidth}
  \centering
  \includegraphics[width=.9\linewidth]{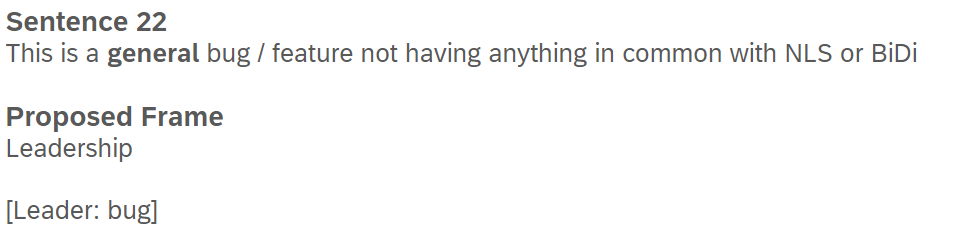}
  \caption{Incorrect frame}
  \label{fig:incorrect}
\end{subfigure}
\caption{Validation frames. The frame in Figure~\ref{fig:correct} is a correct representation of the sentence,
while the frame in Figure~\ref{fig:incorrect} is clearly wrong}
\label{fig:study2validation}
\end{figure}

We then recruited 20 evaluators unknown to us
through Amazon Mechanical
Turk. These evaluators had to have obtained 
the Master
Qualification after repeated participation in the platform.
With this approach, we did not have direct control over the selection of evaluators. We introduced 
two additional sentences in those seen
by an evaluator to validate the evaluator's responses and to avoid exploitation of the platform
with the goal of remuneration. The two validation sentences were hand-picked by the authors.
The first one had an obviously correct frame associated with (Figure~\ref{fig:correct}), while the other
had an obviously wrong frame(Figure~\ref{fig:incorrect}).
The correct one was placed as the first sentence in the study, while the incorrect one was placed at the end.
Neither of these frames were considered in the results.
This validation process lead to two evaluators being replaced, as they had indicated every frame, including
the validation ones, as correct.
With this validation step, we contend that
the evaluators in this study
are a reasonable
representation of novice individuals doing program comprehension; novices may be more likely to
benefit from tools built using technology like \api{}.

Before signing up for the study, evaluators could see how much they would be remunerated and a short description of
the study. Once they had signed up, they received a link to our study
and instructions on how to obtain a unique code to verify they had completed the study.

\red{When starting
the study, evaluators were provided a more detailed description of the task,
including instruction on how to complete the study. They were then presented a list of sentences
with the associated frame, a definition of the frame and any frame elements
the frame might have had.} An example of this can be seen in Figure~\ref{fig:study2validation}.
Participants were asked to indicate for each sentence if the frame correctly represents the sentence.

\subsection{Results}

Table~\ref{tab:study2} summarizes the results of our study. Overall,
in 296 cases (74\%) the participants indicated that the frame was
correct for the sentence. In the majority of cases, 13 out of 20
frames, the majority of participants (at least 3/4 of them) indicated
that the frame extracted was correct. In only one case, 
\textit{Sufficiency}, the majority of participants disagreed with the
proposed frame, but 9 out of 20 still marked it as correct. 

Consistent results from a different set of evaluators on different
datasets increase our confidence that our 
approach generalizes to a broad range of text that appears in software engineering
artifacts.

\section{Threats}
\label{sec:threats}
We performed three investigations as part of exploring the applicability of
semantic frames to 
text associated with program comprehension activities.
Although the details of each
investigation vary, there are similar threats to validity across
the studies.

\subsubsection{Internal and Construct Validity}

We developed~\api{} that tailors semantic frames for software engineering
based on data sampled from three datasets described in Table~\ref{tbl:datasets}.
As this dataset is not representative of all kinds of text appearing
in software engineering artifacts associated
with program comprehension activities,
it is possible that~\api{}
may not adequately parse text from artifacts not considered
by our investigation. 
To address this threat, we included an additional dataset when we
investigated the validity of~\api{} (Section~\ref{sec:study-I}) and
then used a broader set of artifact types and new data---Table~\ref{tbl:datasets-2}---in our
investigation of the generalizability of~\api{}
(Section~\ref{sec:study-II}).

\begin{table}
\centering
\caption{Results of Robustness Study}
{\renewcommand{\arraystretch}{1.05}
\rowcolors{2}{}{lightgray}
\begin{tabular}{lccc}
\hline
\textbf{Frame} & \textbf{Correct} & \textbf{Incorrect} & \textbf{Ratio}  \\
\hline
\hline
Likelihood & 20 & 0 & 100\%  \\
Required Event & 18 & 2 & 90\%  \\
Reasoning & 17 & 3 & 85\%  \\
Existence & 17 & 3 & 85\% \\
Intentionally Act & 17 & 3 & 85\% \\
Relative Time & 17 & 3 & 85\% \\
Time Vector & 17 & 3 & 85\% \\
Events & 16 & 4 & 80\% \\
Sole Instance & 16 & 4 & 80\% \\
Capability & 16 & 4 & 80\% \\
Quantity & 15 & 5 & 75\% \\
Using & 15 & 5 & 75\% \\
Execution & 15 & 5 & 75\% \\
Inclusion & 13 & 7 & 65\% \\
Similarity & 13 & 7 & 65\% \\
Increment & 12 & 8 & 60\% \\
Aggregate & 12 & 8 & 60\% \\
Causation & 11 & 9 & 55\% \\
Temporal Collocation & 10 & 10 & 50\% \\
Sufficiency & 9 & 11 & 45\% \\
\hline
\hline
\textbf{Overall} & 296 & 104 & 74\% \\
\hline
\end{tabular}
}
\label{tab:study2}
\end{table}

Another threat lies in the sentences sampled from the
datasets, which may not be representative. 
To strike a balance between enough examples drawn from 
a diverse number of frames, we considered
the distribution of frames over sampled sentences
such that the focus was on
the frames most likely to appear if~\api{} was applied.
This led to 
the selection of the top 50 (Section~\ref{sec:coding}) and top 20
(Section~\ref{sec:study-II}) most occurring frames in the datasets. To ensure we
were not biased in the artifacts from which we selected sentences,
we employed a sampling approach where we computed the frames for each sentence in every dataset,
and randomly sampled out of these, regardless of which artifact they belong.
This methodology allowed us to 
focus on the topmost occurring frames regardless of how often they appear in the data. 

A key aim in our investigations is to determine the validity of frames
parsed for a sentence. We relied on a variety of evaluators
for the determination of validity. Across the investigation of validity
and robustness, 30 individuals, none of whom are authors of this paper,
assessed frames and sentences. While 1/3 of these evaluators were known
to the authors (Section~\ref{sec:study-II}), 2/3 were not. The use of a broad range
of individuals at arms-length from the authors decreases potential bias
and increases our confidence in the results.
A threat does arise from our choice of not requiring our evaluators recruited through
the MTurk platform to have some certification of English fluency. Recent investigations
in the demographic of the MTurk population has shown that the large majority of MTurkers
is from the USA (75\%)~\cite{difallah2018demographics}. Additionally, our description of the study
was provided in English, and we argue that it is unlikely that someone could have understood the task and
passed the validation questions without a good understanding of the language.

\red{
There are also threats related to the design of \api{} itself.
In Section~\ref{sec:api},
a common pattern that causes a frame to be incorrect for software engineering text is
when source-code related nouns in a sentence evoke a frame.
Instead of creating a list with source-code related terms that \api{} could ignore, 
we opt to discard the frame entirely. 
This design decision errs on the side of caution
because creating and ensuring that a list of source-code related terms 
is complete and always up-to-date is challenging. 
However, this means losing frames that are potentially 
correct in certain cases.
}

\subsubsection{External Validity}

In Section~\ref{sec:study-II}, we investigate the generalizability of our study~\cite{Siegmund2015}. To achieve this goal, we selected 5 new sources of text to use in place of the datasets used in the previous study. Based on 
our methodology and the diverse set of data we used, we believe our results are generalizable to different kinds of text in the domain.

\section{Discussion}
\label{sec:discussion}

Software engineering researchers have recognized limitations of
natural language techniques based on lexicon analysis for processing
software engineering text.  To overcome these limitations, some
software engineering researchers have started to make use of semantic
frame parsing. We have shown that many semantic frames apply to text
appearing in a large range of software artifacts associated with program comprehension activities and that minor
modifications can help address differences in use of language in
software engineering text. However, questions remain about whether
semantic frames can help disambiguate software engineering text where
other approaches struggle and whether our modifications address
problems seen with text in software engineering applications. In this
section, we explore these questions and how they might be addressed in
future work.

\subsection{Using Semantic Frames to Disambiguate SE Text}
\label{sec:apply}

There are limitations associated with applying
lexicon analysis to text associated with program comprehension and
software engineering activities as outlined
in Section~\ref{sec:background}.
To overcome these limitations,  researchers have
used manual labelling combined with classification approaches.
In defining these classification approaches, researchers rely on
discourse patterns in the form of key phrases and generic placeholders
to infer the \textit{meaning} of software engineering text.  For
example, Di Sorbo et al. define a pattern---``\textit{[someone] should add
  [something]}''---to recognize sentences that propose a solution to a
problem~\cite{DiSorbo2015}.  While these approaches produce acceptable
results, they lag in performance because a pattern-based approach is prone to
misclassification~\cite{Huang2018}.

A common misclassification error occurs due to the presence of modal
verbs---an auxiliary verb that
expresses a possibility, likelihood, suggestions, etc.---in a
sentence. In Di Sorbo et al.'s pattern-based approach for
mining intentions from development emails, the presence of
the modal verb `would' in the sentence---``\textit{One way would be to
  add them in an \#ifdef Q\_QDOC block and document them}''---causes
a misclassification to \red{a  
\textit{feature request}
instead of a \textit{solution proposal}}~\cite{Huang2018}.

While a sentence related to feature request often contain 
 users' desires or aspirations, 
a sentence describing a solution proposal should contain directives or instructions; this meaning would
be captured by the \textit{Execution} frame in semantic frame parsing.
Unfortunately, generic frame parsing does not suggest
such a frame.  However, \api{}, when applied to this sentence, 
identifies such a frame what could have assisted correctly classifying Di Sorbo et al.'s example.

In another piece of software engineering research, Chapparo et
al. attempt to classify if a sentence of a bug report contains
`observed behaviour', `expected behaviour', or `steps to reproduce the
bug'~\cite{Chaparro2017}.  Similar to Di Sorbo, Chapparo et al. also
observe misclassification errors.  For example, the presence of a
modal verb causes their approach to misclassify a sentence discussing
an observed behaviour ``\textit{this problem could also be related to
  some sites not copying URLs}'' as an expected behaviour.  For this
sentence, \api{} parses the \textit{Predicament} frame, which is
a frame that evaluators consistently agreed is
related to undesired behaviours.  The presence of this frame, as
identified by \api, could have assisted determining that the sentence
is about a bug's observed behaviour.

We suggest that future work  
consider how classifiers based on \api{} compares to pattern matching-based
classifiers (e.g.,~\cite{DiSorbo2015,Chaparro2017,Robillard2015}).

\subsection{Use and Limitations of SEFrame}
\label{sec:improvement}

Software engineering researchers have noted the need to interpret text
in a structured way. For example, Robillard et al. show that the
structure of information in API documentation can help facilitate
discussions about the content of APIs~\cite{Maalej2013,
  Robillard2015}. The structure extracted from a sentence by
frame semantics could help in these situations and in particular,
\api{} may help with extracting structure from software engineering text
in applications such as those described by Robillard et al.
Consider
an example
from Robillard and Chhetri~\cite{Robillard2015} where a sentence contains a `\textit{directive}'
for cloning objects in the Java API 
``\textit{By convention, the returned object should be obtained by calling super.clone}''.
Figure~\ref{fig:frame-design} shows the same sentence in a structured format
what suggests obligations, or requirements, for proper usage of the clone method.

\begin{figure}[h!]
\medskip
\centering
\begin{small}
\begingroup
\setlength{\tabcolsep}{10pt} \renewcommand{\arraystretch}{1.2} \centering  
\begin{tabular}{lcl}

\hline

& & 
By convention,
\\

{\ttfamily \scriptsize \color{rufous} \textbf{fe:Responsible party}} 
& $\rightarrow$ &
the returned object 
\\

{\ttfamily \scriptsize \hl{\textbf{Being obligated}}} 
& $\rightarrow$ &
should 
\\

{\ttfamily \scriptsize \color{rufous} \textbf{fe:Duty} } 
& $\rightarrow$ &
be obtained
\\

& &
by 
\\

{\ttfamily \scriptsize \hl{\textbf{Execution}}} 
& $\rightarrow$ &
calling
\\

{\ttfamily \scriptsize \color{rufous} \textbf{fe:Target} } 
& $\rightarrow$ &
super.clone
\\

\hline

\end{tabular}
\endgroup
\end{small}
\caption{Example of structured information for an API directive from ~\cite{Robillard2015}}
\label{fig:frame-design}
\end{figure}

We have assessed and tailored frame semantics across a broad range
of software engineering text associated with program comprehension. Upon application of \api{} to
program comprehension tasks, it may be found that further
improvements are needed by analyzing more types of frames
or by adopting more complex semantic parsers (e.g.,~\cite{roth2015context, fitzgerald2015semantic} or ~\cite{swayamdipta2017frame}).
 It is likely this analysis will
require a similar approach to what we have taken in this paper that includes manual
inspection and annotation.  Future studies can benefit from the
sampling methodology and annotation framework we outline to aid in these
extensions. The frames on which to focus may be guided by a larger analysis
of the distribution of frame occurrence over a wider set of software
engineering artifacts.

\section{Conclusion}
\label{sec:conclusion}

Software engineering researchers have noted that general natural language processing
techniques based on lexicon analysis are not sufficient for all software engineering
problems. In this paper, we have investigated whether frame semantics
might help fill this gap 
for a range of software artifacts associated
with program comprehension activities.
Frame semantics is a general linguistic
approach that extracts semantic frames from natural language
sentences, aiming to represent the intended meaning of the words in a
sentence.

We developed a tool, called \api, that tailors an
every-day English semantic frame parser (SEMAFOR) 
for 
artifacts associated with program comprehension
activities. We
evaluated \api{} in two ways: first to evaluate its correctness
(Section~\ref{sec:study-I}) and then to evaluate its robustness
(Section~\ref{sec:study-II}). In both studies, \api{} performed well,
extracting the correct frame in 73\% of the cases in the first study
and 74\% in the second study. Based on these results, we argue that
\api{} can reliably extract the correct frame for sentences across
a broad range of software engineering text
appearing in artifacts used to support
program comprehension.
These frames could be used to
successfully categorize semantic information across a variety of
artifacts commonly sought during software development.

We plan to continue expanding \api{} to include an
evaluation of all the frames presents in the FrameNet project to more
accurately classify the semantic information in software engineering
text and to explore its application to software engineering problems.

\section*{Acknowledgements}
The authors would like to thank the participants of the correctness and robustness studies, and the anonymous reviewers for their valuable feedback.
We also thank Shaunak Tulshibagwale for his help implementing some of the datasets samplers.

\bibliographystyle{IEEEtran}
\bibliography{paper}
\end{document}